\documentclass[%
reprint,
showpacs,preprintnumbers,
nofootinbib,
amsmath,amssymb,
aps,
prd,
floatfix,
]{revtex4-1}

\usepackage{graphicx}% Include figure files
\usepackage{dcolumn}% Align table columns on decimal point
\usepackage{bm}% bold math

\usepackage{url}

\newcommand{\boss}[2]{\ensuremath{\rlap{\kern-2.5pt\ensuremath{\overset{\scriptscriptstyle(-)}{\phantom{#1}}}}{\ensuremath{{#1}_{#2}}}}}

\begin{document}

\preprint{\begin{tabular}{l}
\texttt{EURONU-WP6-11-37}
\\
\texttt{arXiv:1107.1452v3 [hep-ph]}
\end{tabular}}

\title{3+1 and 3+2 Sterile Neutrino Fits}

\author{Carlo Giunti}
\email{giunti@to.infn.it}
\altaffiliation[also at ]{Department of Theoretical Physics, University of Torino, Italy}
\affiliation{INFN, Sezione di Torino, Via P. Giuria 1, I--10125 Torino, Italy}

\author{Marco Laveder}
\email{laveder@pd.infn.it}
\affiliation{Dipartimento di Fisica ``G. Galilei'', Universit\`a di Padova,
and
INFN, Sezione di Padova,
Via F. Marzolo 8, I--35131 Padova, Italy}

\date{\today}

\begin{abstract}
We present the results of fits of short-baseline neutrino oscillation data
in 3+1 and 3+2 neutrino mixing schemes.
In spite of the presence of a tension in the interpretation of the data,
3+1 neutrino mixing is attractive for its simplicity and for the natural correspondence
of one new entity
(a sterile neutrino)
with a new effect
(short-baseline oscillations).
The allowed regions in the oscillation parameter space can be tested in near-future experiments.
In the framework of 3+2 neutrino mixing there is less tension in the interpretation of the data,
at the price of introducing a second sterile neutrino.
Moreover,
the improvement of the parameter goodness of fit is mainly a statistical effect due to an increase of
the number of parameters.
The CP violation in short-baseline experiments allowed in 3+2 neutrino mixing
can explain the positive $\bar\nu_{\mu}\to\bar\nu_{e}$ signal
and
the negative $\nu_{\mu}\to\nu_{e}$ measurement
in the MiniBooNE experiment.
For the CP-violating phase we obtained
two minima of the marginal $\chi^2$ close to the two values where CP-violation is maximal.
\end{abstract}


\maketitle

\section{Introduction}

The recent agreement of
MiniBooNE antineutrino data
\cite{1007.1150}
with the short-baseline $\bar\nu_{\mu}\to\bar\nu_{e}$ oscillation signal
observed several years ago in the LSND experiment
\cite{hep-ex/0104049}
has opened an intense theoretical and experimental
activity aimed at the clarification of the
explanation of these observations
in a framework compatible with the data of other neutrino oscillation experiments.
Several short-baseline neutrino oscillation experiments did not observe
neutrino oscillations and their data constraint the interpretation of the
LSND and MiniBooNE antineutrino signal.
However,
there are other positive indications of short-baseline neutrino oscillations that may be taken into account:
the reactor antineutrino anomaly
\cite{1101.2755},
in favor of a small short-baseline disappearance of $\bar\nu_{e}$,
the Gallium neutrino anomaly
\cite{hep-ph/9411414,Laveder:2007zz,hep-ph/0610352,0707.4593,0711.4222,0902.1992,1005.4599,1006.2103,1006.3244,1101.2755},
in favor of a short-baseline disappearance of $\nu_{e}$,
and
the MiniBooNE low-energy anomaly \cite{0707.4593,0902.1992,1005.4599,1101.2755}.
In this paper we consider only the reactor antineutrino anomaly,
by taking into account the new calculation of reactor antineutrino fluxes in Ref.~\cite{1101.2663}.
We leave the discussion of the effects of the more controversial
Gallium anomaly and MiniBooNE low-energy anomaly
to a following article
\cite{Giunti-Laveder-IP-11}.

The results of
solar, atmospheric and long-baseline neutrino oscillation experiments
led us to the current standard three-neutrino mixing paradigm,
in which the three active neutrinos
$\nu_{e}$,
$\nu_{\mu}$,
$\nu_{\tau}$
are superpositions of three massive neutrinos
$\nu_1$,
$\nu_2$,
$\nu_3$
with respective masses
$m_1$,
$m_2$,
$m_3$.
The measured solar (SOL) and atmospheric (ATM) squared-mass differences can be interpreted as
\begin{align}
\null & \null
\Delta{m}^2_{\text{SOL}}
=
\Delta{m}^2_{21}
=
(7.6 \pm 0.2) \times 10^{-5} \, \text{eV}^2
\quad
\text{\protect\cite{1010.0118}}
\,,
\label{SOL}
\\
\null & \null
\Delta{m}^2_{\text{ATM}}
=
|\Delta{m}^2_{31}|
=
2.32 {}^{+0.12}_{-0.08} \times 10^{-3} \, \text{eV}^2
\quad
\text{\protect\cite{1103.0340}}
\,,
\label{ATM}
\end{align}
with
$\Delta{m}^2_{kj}=m_k^2-m_j^2$.

The completeness of the three-neutrino mixing paradigm has been
challenged by the
LSND \cite{hep-ex/0104049} and MiniBooNE \cite{1007.1150}
observations of short-baseline
$\bar\nu_{\mu}\to\bar\nu_{e}$ transitions
at
different values of distance ($L$) and energy ($E$),
but approximately at the same $L/E$.
Since the distance and energy dependences of
neutrino oscillations occur through this ratio,
the agreement of the MiniBooNE and LSND signals raised
interest in the possibility of existence of
one or more squared-mass differences larger than about 0.5 eV,
which is much bigger than the values of
$\Delta{m}^2_{\text{SOL}}$
and
$\Delta{m}^2_{\text{ATM}}$.
Hence,
we are lead to the extension of three-neutrino mixing with the introduction of one or more sterile neutrinos
which do not have weak interactions and do not contribute
to the invisible width of the $Z$ boson
\cite{hep-ex/0509008}.
In this paper we consider the simplest possibilities:
3+1 mixing with one sterile neutrino
and
3+2 mixing with two sterile neutrinos.

The existence of sterile neutrinos
which have been thermalized in the early Universe is compatible
with Big-Bang Nucleosynthesis data
\cite{astro-ph/0408033,1001.4440},
with the indication however that schemes with more than one sterile neutrino are disfavored
\cite{1103.1261},
and cosmological measurements of the
Cosmic Microwave Background and Large-Scale Structures
if the neutrino masses are limited below about 1 eV
\cite{1006.5276,1102.4774,1104.0704,1104.2333,1106.5052}.
Therefore,
in this paper we consider squared-mass differences smaller than $10 \, \text{eV}^2$.

\begin{figure}[t!]
\null
\hfill
\begin{minipage}[l]{0.2\linewidth}
\begin{center}
\includegraphics*[bb=236 471 342 778, width=0.9\linewidth]{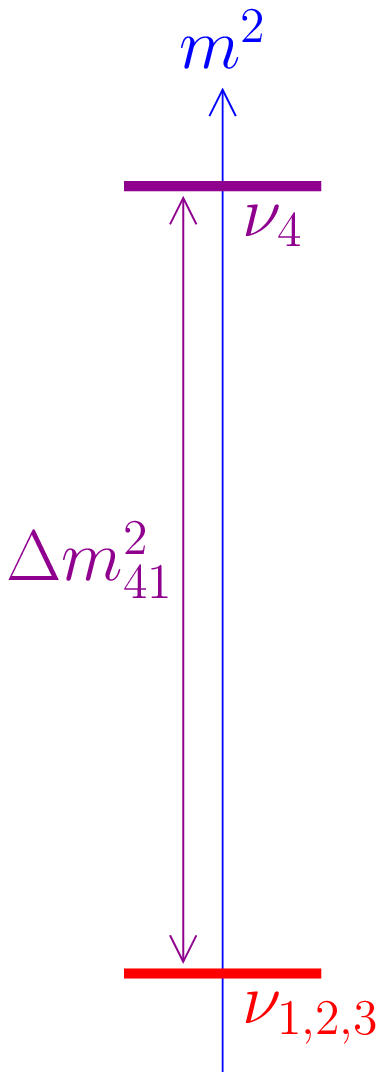}
\\%[0.3cm]
\text{{"normal"}}
\end{center}
\end{minipage}
\hfill
\begin{minipage}[l]{0.2\linewidth}
\begin{center}
\includegraphics*[bb=236 471 342 778, width=0.9\linewidth]{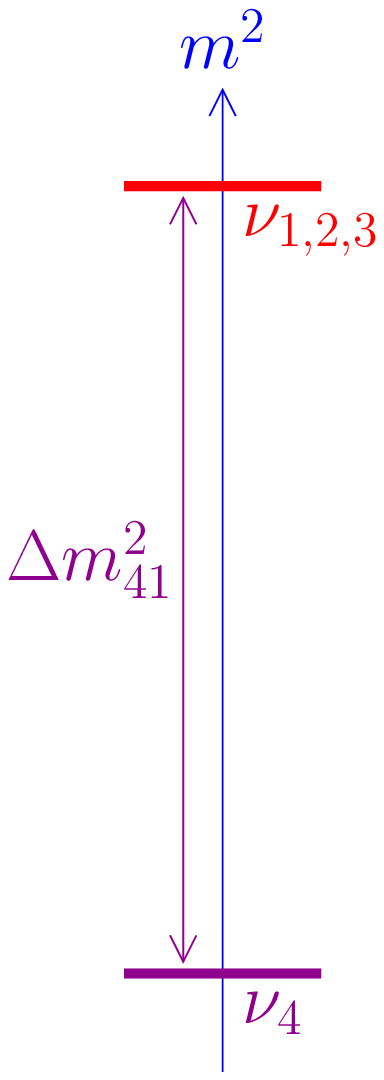}
\\%[0.3cm]
\text{{"inverted"}}
\end{center}
\end{minipage}
\hfill
\null
\caption{ \label{3+1}
Schematic description of the two possible
3+1 schemes that we are considering, taking into account that
$|\Delta{m}^2_{21}| \ll |\Delta{m}^2_{31}| \ll |\Delta{m}^2_{41}|$.
}
\end{figure}

\section{3+1 Neutrino Mixing}

In this section we consider the
simplest extension of three-neutrino mixing
with the addition of one massive neutrino.
In such four-neutrino mixing framework
the flavor neutrino basis is composed by the three active neutrinos
$\nu_{e}$,
$\nu_{\mu}$,
$\nu_{\tau}$
and
a sterile neutrino
$\nu_{s}$.

So-called 2+2 four-neutrino mixing schemes are strongly disfavored
by the absence of any signal of sterile neutrino effects in
solar and atmospheric neutrino data
\cite{hep-ph/0405172}.
Hence, we consider only the so-called 3+1 four-neutrino schemes
depicted in Fig.~\ref{3+1}.
Since the
inverted
scheme has three massive neutrinos at the eV scale,
it is disfavored by cosmological data \cite{1006.5276,1102.4774}
over the
normal scheme.
In both 3+1 schemes the
effective flavor transition and survival probabilities
in short-baseline (SBL) experiments
are given by
\begin{align}
\null & \null
P_{\boss{\nu}{\alpha}\to\boss{\nu}{\beta}}^{\text{SBL}}
=
\sin^{2} 2\vartheta_{\alpha\beta}
\sin^{2}\left( \frac{\Delta{m}^2_{41} L}{4E} \right)
\qquad
(\alpha\neq\beta)
\,,
\label{trans}
\\
\null & \null
P_{\boss{\nu}{\alpha}\to\boss{\nu}{\alpha}}^{\text{SBL}}
=
1
-
\sin^{2} 2\vartheta_{\alpha\alpha}
\sin^{2}\left( \frac{\Delta{m}^2_{41} L}{4E} \right)
\,,
\label{survi}
\end{align}
for
$\alpha,\beta=e,\mu,\tau,s$,
with
\begin{align}
\null & \null
\sin^{2} 2\vartheta_{\alpha\beta}
=
4 |U_{\alpha4}|^2 |U_{\beta4}|^2
\,,
\label{transsin}
\\
\null & \null
\sin^{2} 2\vartheta_{\alpha\alpha}
=
4 |U_{\alpha4}|^2 \left( 1 - |U_{\alpha4}|^2 \right)
\,.
\label{survisin}
\end{align}
Therefore:
\begin{enumerate}
\item
All effective SBL oscillation probabilities
depend only on the absolute value of the largest squared-mass difference
$\Delta{m}^2_{41}$.
\item
All oscillation channels are open, each one with its own oscillation amplitude.
\item
The oscillation amplitudes depend only on the absolute values
of the elements in the fourth column of the mixing matrix,
i.e. on three real numbers with sum less than unity,
since the unitarity of the mixing matrix implies
$
\sum_{\alpha} |U_{\alpha4}|^2 = 1
$
\item
CP violation cannot be observed in SBL oscillation experiments,
even if the mixing matrix contains CP-violation phases.
In other words,
neutrinos and antineutrinos have the same
effective SBL oscillation probabilities.
\end{enumerate}

Before the recent indication of an antineutrino
$\bar\nu_{\mu}\to\bar\nu_{e}$
signal consistent with the
LSND antineutrino signal,
the MiniBooNE collaboration published
the results of neutrino data which do not show
a corresponding $\nu_{\mu}\to\nu_{e}$
signal
\cite{0812.2243}.
This difference between the MiniBooNE neutrino and antineutrino data may be due to CP violation.

The absence of any difference in the
effective SBL oscillation probabilities
of
neutrinos and antineutrinos
in 3+1 four-neutrino mixing schemes
implies that these schemes
cannot explain the difference between
neutrinos and antineutrino oscillations observed in the
MiniBooNE experiment.
Moreover,
the dependence of all the oscillation amplitudes in Eqs.~(\ref{transsin}) and (\ref{survisin})
on three independent absolute values
of the elements in the fourth column of the mixing matrix
implies that the amplitude of
$
\boss{\nu}{\mu}\to\boss{\nu}{e}
$
transitions is limited by the absence of large SBL disappearance of
$\boss{\nu}{e}$ and $\boss{\nu}{\mu}$
observed in several experiments.

\begin{figure}[t!]
\includegraphics*[width=\linewidth]{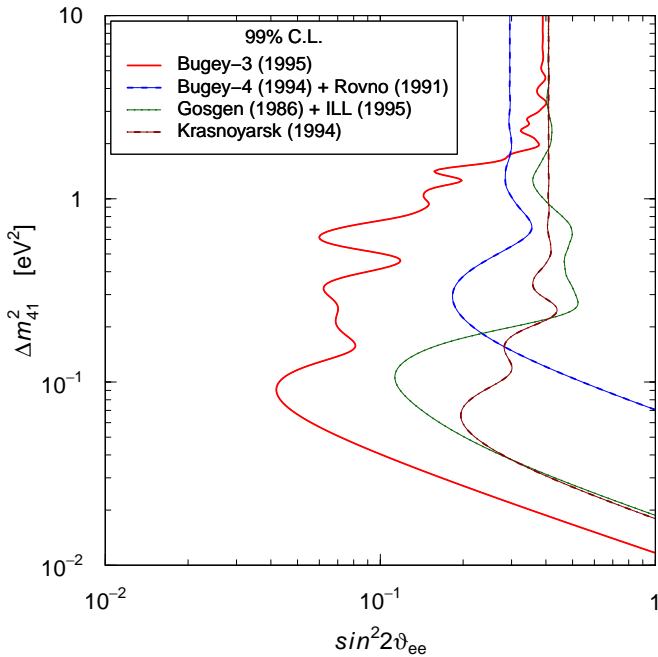}
\caption{ \label{rea-cnt}
Exclusion curves obtained from the data of reactor $\bar\nu_{e}$ disappearance experiments
(see Ref.~\cite{1101.2755}).
}
\end{figure}

The results of reactor neutrino experiments constrain the value
$|U_{e4}|^2$
through the measurement of
$\sin^{2} 2\vartheta_{ee}$.
The calculation of the reactor $\bar\nu_{e}$ flux has been recently improved in Ref.~\cite{1101.2663},
resulting in an increase of about 3\% with respect to the previous value adopted
by all experiments for the comparison with the data
(see Ref.~\cite{hep-ph/0107277}).
Since the measured reactor rates are in approximate agreement with those derived from the old $\bar\nu_{e}$ flux,
they show a deficit with respect to the rates derived from the new $\bar\nu_{e}$ flux.
This is the ``reactor antineutrino anomaly'' \cite{1101.2755},
which is quantified by the value
\begin{equation}
\overline{R}_{\text{reactor anomaly}}
=
0.946 \pm 0.024
\label{reactoranomaly}
\end{equation}
for the average of the ratios of measured event rates and those expected in absence of $\bar\nu_{e}$ transformations into other states.
We considered the integral rates of the
Bugey-3 \cite{Declais:1995su},
Bugey-4 \cite{Declais:1994ma},
ROVNO91 \cite{Kuvshinnikov:1990ry},
Gosgen \cite{Zacek:1986cu}
and
Krasnoyarsk \cite{Vidyakin:1990iz}
short-baseline reactor antineutrino experiments
using the information in Table~II of Ref.~\cite{1101.2755}\footnote{
We do not use the two rates of the
Savannah River experiment \cite{Greenwood:1996pb}
in Table~II of Ref.~\cite{1101.2755},
$R_{\text{SRP I}} = 0.952 \pm 0.006 \pm 0.037$
and
$R_{\text{SRP II}} = 1.018 \pm 0.010 \pm 0.037$
because they are about $5.5\sigma$ apart,
taking into account that their difference
$0.066 \pm 0.012$
is independent of the correlated systematic uncertainty (0.037).
Such a large difference cannot be due to neutrino oscillations averaged over the whole energy spectrum,
because the two measurements have been done at distances which are not different enough
(18 m and 24 m).
We also do not use the ROVNO88 \cite{Afonin:1988gx} rates
in Table~II of Ref.~\cite{1101.2755},
because the correlation with Bugey-4 and ROVNO91
is not clear.
}.
Hence, the reactor antineutrino anomaly is a $2.2\sigma$
indication that there is a small short-baseline disappearance of $\bar\nu_{e}$ 
which may correspond to the
$\bar\nu_{\mu}\to\bar\nu_{e}$
signal observed in the
LSND and MiniBooNE experiments.
However,
the $\bar\nu_{e}$ disappearance is small and large values of
$\sin^{2} 2\vartheta_{ee}$
are constrained by the exclusion curves in Fig.~\ref{rea-cnt}
(as in Ref.~\cite{1101.2755},
the Bugey-3 exclusion curve has been obtained by fitting the three integral rates measured
at $L=15,40,95\,\text{m}$ and the $40\,\text{m}/15\,\text{m}$ spectral ratio in Fig.~15 of Ref.~\cite{Declais:1995su}).
Since values of $|U_{e4}|^2$ close to unity are excluded by solar neutrino oscillations
(which require large $|U_{e1}|^2+|U_{e2}|^2$),
for small $\sin^{2} 2\vartheta_{ee}$ we have
\begin{equation}
\sin^{2} 2\vartheta_{ee} \simeq 4 |U_{e4}|^2
\,.
\label{ue4}
\end{equation}

\begin{figure}[t!]
\includegraphics*[width=\linewidth]{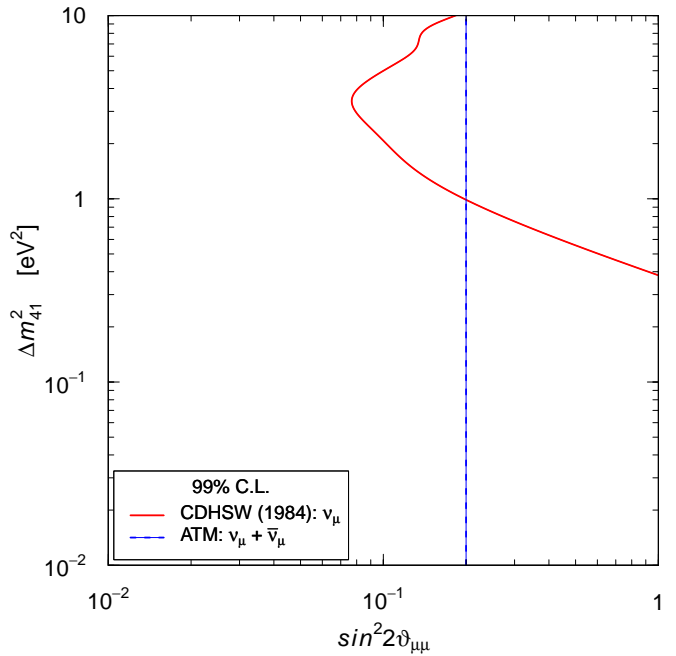}
\caption{ \label{dmu-sup}
Exclusion curves obtained
from the data of the CDHSW $\nu_{\mu}$ disappearance experiment
\cite{Dydak:1984zq},
and from atmospheric neutrino data
(extracted from the analysis in Ref.~\cite{0705.0107}).
}
\end{figure}

The value of $\sin^{2} 2\vartheta_{\mu\mu}$
is constrained by the curves in Fig.~\ref{dmu-sup},
which have been obtained from
the lack of $\nu_{\mu}$ disappearance in the CDHSW $\nu_{\mu}$ experiment
\cite{Dydak:1984zq}
and
from the requirement of large $|U_{\mu1}|^2+|U_{\mu2}|^2+|U_{\mu3}|^2$
for atmospheric neutrino oscillations \cite{0705.0107}.
Hence,
$|U_{\mu4}|^2$ is small and
\begin{equation}
\sin^{2} 2\vartheta_{\mu\mu} \simeq 4 |U_{\mu4}|^2
\,.
\label{um4}
\end{equation}

\begin{figure}[t!]
\includegraphics*[width=\linewidth]{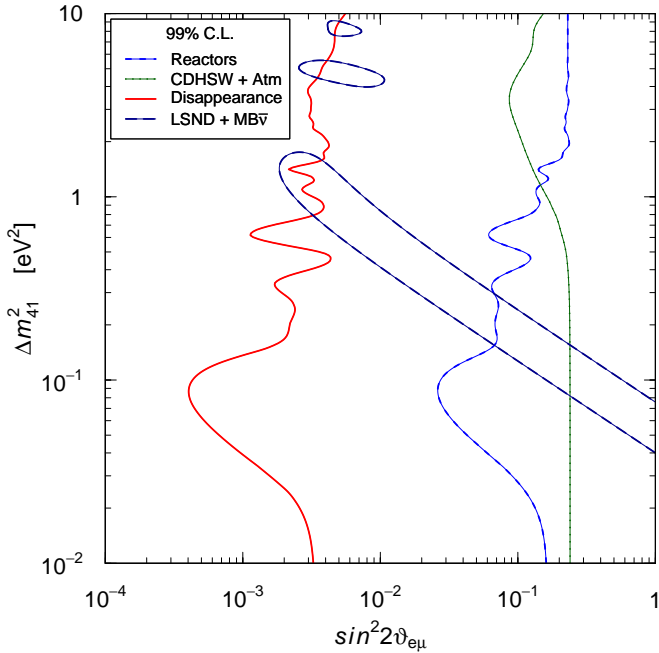}
\caption{ \label{exc-dis}
Exclusion curves
in the $\sin^{2} 2\vartheta_{e\mu}$--$\Delta{m}^2_{41}$ plane
obtained from the separate constraints in Figs.~\ref{rea-cnt} and \ref{dmu-sup}
(blue dashed line and green dotted line)
and the combined constraint given by Eq.~(\ref{sem-exc})
(red solid line)
from disappearance experiments (Dis).
The regions allowed by LSND and MiniBooNE antineutrino data are delimited by
dark-blue long-dashed lines.
}
\end{figure}

\begin{figure}[t!]
\includegraphics*[width=\linewidth]{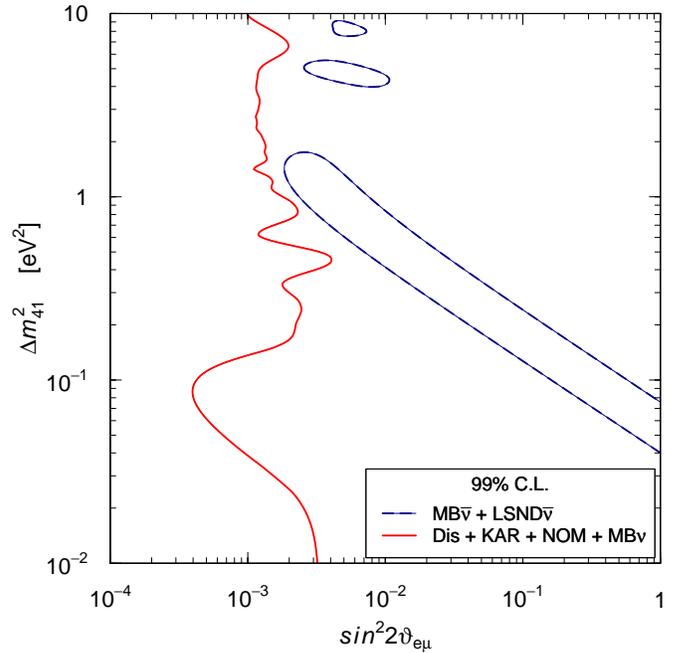}
\caption{ \label{exc-nev}
Exclusion curve
in the $\sin^{2} 2\vartheta_{e\mu}$--$\Delta{m}^2_{41}$ plane
obtained with the addition to the
disappearance constraint in Fig.~\ref{exc-dis} of the constraints obtained from
KARMEN \cite{hep-ex/0203021} (KAR),
NOMAD \cite{hep-ex/0306037} (NOM)
and 
MiniBooNE neutrino \cite{0812.2243} (MB$\nu$)
data (red solid line).
The regions allowed by LSND and MiniBooNE antineutrino data are delimited by
dark-blue long-dashed lines.
}
\end{figure}

From Eqs.~(\ref{transsin}), (\ref{ue4}) and (\ref{um4}),
for the amplitude of
$
\boss{\nu}{\mu}\to\boss{\nu}{e}
$
transitions we obtain
\cite{hep-ph/9606411,hep-ph/9607372}
\begin{equation}
\sin^{2} 2\vartheta_{e\mu}
\simeq
\frac{1}{4}
\,
\sin^{2} 2\vartheta_{ee}
\,
\sin^{2} 2\vartheta_{\mu\mu}
\,.
\label{sem-exc}
\end{equation}
Therefore,
if
$\sin^{2} 2\vartheta_{ee}$
and
$\sin^{2} 2\vartheta_{\mu\mu}$
are small,
$\sin^{2} 2\vartheta_{e\mu}$
is quadratically suppressed.
This is illustrated in Fig.~\ref{exc-dis},
where one can see that the separate effects of the constraints on
$\sin^{2} 2\vartheta_{ee}$
and
$\sin^{2} 2\vartheta_{\mu\mu}$
exclude only the large-$\sin^{2} 2\vartheta_{e\mu}$
part of the region allowed by
LSND and MiniBooNE antineutrino data,
whereas most of this region is excluded by the combined constraint in Eq.~(\ref{sem-exc}).
As shown in Fig.~\ref{exc-nev},
the constraint becomes stronger by including the data of the
KARMEN \cite{hep-ex/0203021},
NOMAD \cite{hep-ex/0306037}
and 
MiniBooNE neutrino \cite{0812.2243}
experiments,
which did not observe a short-baseline
$
\boss{\nu}{\mu}\to\boss{\nu}{e}
$
signal.
Since the parameter goodness-of-fit
\cite{hep-ph/0304176}
is
$
6
\times
10^{-6}
$,
3+1 schemes are disfavored by the data.
This conclusion has been reached recently also in Refs.~\cite{0705.0107,1007.4171,1012.0267,1103.4570}
and confirms the pre-MiniBooNE results in Refs.~\cite{hep-ph/9607372,hep-ph/9903454,hep-ph/0207157,hep-ph/0405172}.

However,
in spite of the low value of the parameter goodness-of-fit
it is not inconceivable to refuse to reject the 3+1 schemes
for the following reasons:
\begin{enumerate}
\item
It is the simplest scheme beyond the standard three-neutrino mixing which can partially explain the data.
\item
It corresponds to the natural addition of one new entity (a sterile neutrino) to explain a new effect
(short-baseline oscillations).
Better fits of the data require the addition of at least another new entity
(in any case at least one sterile neutrino is needed to generate short-baseline oscillations).
\item
The minimum value of the global $\chi^2$ is rather good:
$\chi^2_{\text{min}} = 100.2$
for
$104$
degrees of freedom.
%corresponding to a goodness of fit of
%$59\%$.
\end{enumerate}
Note also that
3+1 schemes are favored with respect to 3+2 schemes by the
Big-Bang Nucleosynthesis limit
$N_{\text{eff}}\leq4$ at 95\% C.L.
obtained in Ref.~\cite{1103.1261}.
Therefore,
we consider the global fit of all data in 3+1 schemes,
which yields the best-fit values of the oscillation parameters listed in Tab.~\ref{bef}.
%\begin{equation}
%\Delta{m}^2_{41} = 0.89 \, \text{eV}^2
%\,,\,
%|U_{e4}|^2 = 0.025
%\,,\,
%|U_{\mu4}|^2 = 0.023
%\,.
%\label{fit-3p1}
%\end{equation}

\begin{table}[b!]
\begin{center}
\begin{tabular}{ccc}
&
3+1
&
3+2
\\
\hline
$\chi^2_{\text{min}}$
&
$100.2$
&
$91.6$
\\
$\text{NDF}$
&
$104$
&
$100$
\\
$\text{GoF}$
&
$59\%$
&
$71\%$
\\
\hline
$\Delta{m}^2_{41}\,[\text{eV}^2]$
&
$0.89$
&
$0.90$
\\
$|U_{e4}|^2$
&
$0.025$
&
$0.017$
\\
$
|U_{\mu4}|^2$
&
$0.023$
&
$0.019$
\\
$\Delta{m}^2_{51}\,[\text{eV}^2]$
&
&
$1.61$
\\
$|U_{e5}|^2$
&
&
$0.017$
\\
$
|U_{\mu5}|^2$
&
&
$0.0061$
\\
$\eta$
&
&
$1.51 \pi$
\\
\hline
$\Delta\chi^{2}_{\text{PG}}$
&
$24.1$
&
$22.2$
\\
$\text{NDF}_{\text{PG}}$
&
$2$
&
$5$
\\
$\text{PGoF}$
&
$
6
\times
10^{-6}
$
&
$
5
\times
10^{-4}
$
\\
\hline
\end{tabular}
\end{center}
\caption{ \label{bef}
Values of
$\chi^{2}$,
number of degrees of freedom (NDF),
goodness-of-fit (GoF)
and
best-fit values of the mixing parameters
obtained in our 3+1 and 3+2 fits of short-baseline oscillation data.
The last three lines give the results of the parameter goodness-of-fit test
\cite{hep-ph/0304176}:
$\Delta\chi^{2}_{\text{PG}}$,
number of degrees of freedom ($\text{NDF}_{\text{PG}}$) and
parameter goodness-of-fit (PGoF).
}
\end{table}

\begin{figure}[t!]
\includegraphics*[width=\linewidth]{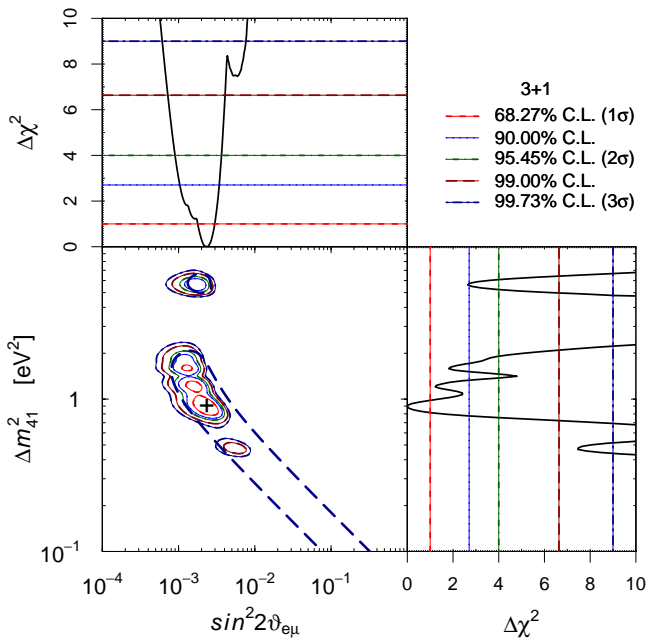}
\caption{ \label{3p1-sem}
Allowed regions in the
$\sin^{2}2\vartheta_{e\mu}$--$\Delta{m}^2_{41}$ plane
and
marginal $\Delta\chi^{2}$'s
for
$\sin^{2}2\vartheta_{e\mu}$ and $\Delta{m}^2_{41}$
obtained from the global fit of all the considered data in 3+1 schemes.
The best-fit point corresponding to $\chi^2_{\text{min}}$ is indicated by a cross.
The isolated dark-blue dash-dotted contours enclose the regions allowed at $3\sigma$
by the analysis of appearance data
(the $\bar\nu_{\mu}\to\bar\nu_{e}$ data of the 
LSND \cite{hep-ex/0104049},
KARMEN \cite{hep-ex/0203021} and
MiniBooNE \cite{1007.1150}
experiments
and
the $\nu_{\mu}\to\nu_{e}$ data of the
NOMAD \cite{hep-ex/0306037} and
MiniBooNE \cite{0812.2243}
experiments).
}
\end{figure}

\begin{figure}[t!]
\includegraphics*[width=\linewidth]{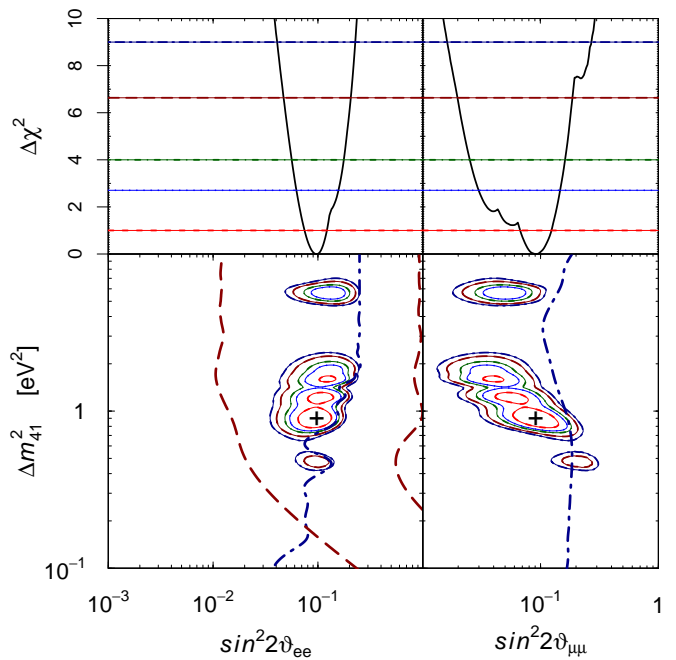}
\caption{ \label{3p1-see-smm}
Allowed regions in the
$\sin^{2}2\vartheta_{ee}$--$\Delta{m}^2_{41}$ and
$\sin^{2}2\vartheta_{\mu\mu}$--$\Delta{m}^2_{41}$ planes
and
marginal $\Delta\chi^{2}$'s
for
$\sin^{2}2\vartheta_{ee}$ and $\sin^{2}2\vartheta_{\mu\mu}$
obtained from the global fit of all the considered data in 3+1 schemes.
The best-fit point corresponding to $\chi^2_{\text{min}}$ is indicated by a cross.
The line types and color have the same meaning as in Fig.~\ref{3p1-sem}.
The isolated dark-blue dash-dotted lines are the $3\sigma$ exclusion curves
obtained from reactor neutrino data in the left plot
(corresponding to the blue dashed line in Fig.~\ref{exc-dis})
and from CDHSW and atmospheric neutrino data in the right plot
(corresponding to the green dotted line in Fig.~\ref{exc-dis}).
The isolated dark-red long-dashed lines delimit the region allowed at 99\% C.L.
by the Gallium anomaly \cite{1006.3244}.
}
\end{figure}

Figures~\ref{3p1-sem} and \ref{3p1-see-smm} show the allowed regions in the
$\sin^{2}2\vartheta_{e\mu}$--$\Delta{m}^2_{41}$,
$\sin^{2}2\vartheta_{ee}$--$\Delta{m}^2_{41}$ and
$\sin^{2}2\vartheta_{\mu\mu}$--$\Delta{m}^2_{41}$ planes
and the marginal $\Delta\chi^{2}$'s
for
$\Delta{m}^2_{41}$,
$\sin^{2}2\vartheta_{e\mu}$,
$\sin^{2}2\vartheta_{ee}$
and
$\sin^{2}2\vartheta_{\mu\mu}$.
The best-fit values of the oscillation amplitudes are
\begin{align}
\null & \null
\sin^{2}2\vartheta_{e\mu}
=
0.0023
\,,
\label{sem}
\\
\null & \null
\sin^{2}2\vartheta_{ee}
=
0.098
\,,
\label{see}
\\
\null & \null
\sin^{2}2\vartheta_{\mu\mu}
=
0.091
\,.
\label{smm}
\end{align}

From Fig.~\ref{3p1-sem} one can see that the allowed regions are compatible with those
allowed by appearance data
(the $\bar\nu_{\mu}\to\bar\nu_{e}$ data of the 
LSND \cite{hep-ex/0104049},
KARMEN \cite{hep-ex/0203021} and
MiniBooNE \cite{1007.1150}
experiments
and
the $\nu_{\mu}\to\nu_{e}$ data of the
NOMAD \cite{hep-ex/0306037} and
MiniBooNE \cite{0812.2243}
experiments)
and are slightly pushed towards the left by the disappearance constraints.
Future experiments aimed at checking the
LSND \cite{hep-ex/0104049}
and
MiniBooNE \cite{1007.1150}
$\bar\nu_{\mu}\to\bar\nu_{e}$ oscillation signal
(as those in Refs.~\cite{CRUBBIA2011,1007.3228,1105.4984})
should aim at exploring these regions.

Figure~\ref{3p1-see-smm} shows that the allowed regions in the
$\sin^{2}2\vartheta_{ee}$--$\Delta{m}^2_{41}$ and
$\sin^{2}2\vartheta_{\mu\mu}$--$\Delta{m}^2_{41}$ planes
lie just on the left of the disappearance constraints,
as expected.
From the left panel in Fig.~\ref{3p1-see-smm}
one can see that the allowed regions in the
$\sin^{2}2\vartheta_{ee}$--$\Delta{m}^2_{41}$
plane are compatible with the area indicated by the
Gallium anomaly \cite{1006.3244}.
The allowed region around the best- fit point
and the isolated region at
$\Delta{m}^2_{41}\simeq6\,\text{eV}^2$
are also compatible with the recent results in Ref.~\cite{1106.5552}.
If the 3+1 neutrino mixing scheme is realized in nature,
future experiments searching for short-baseline
$\boss{\nu}{e}$
disappearance
(as those in Refs.~\cite{0907.3145,0907.5487,1006.2103,Porta:2010zz,1011.4509,1103.5307,CRUBBIA2011,Pallavicini2011,1105.4984})
should find a disappearance compatible with the reactor antineutrino anomaly
in Eq.~(\ref{reactoranomaly}).
Future experiments searching for short-baseline
$\boss{\nu}{\mu}$
disappearance
(as those in Refs.~\cite{CRUBBIA2011,1106.5685})
should find a disappearance just below the current bound,
for
$
0.4
\lesssim
\Delta{m}^2_{41}
\lesssim
7 \, \text{eV}^2
$.
Short-baseline
$\boss{\nu}{e}$
and
$\boss{\nu}{\mu}$
disappearance can have observable effects, respectively, also in solar neutrino experiments
\cite{0910.5856,1105.1705},
long-baseline neutrino oscillation experiments
\cite{1104.3922,1105.5946}
and atmospheric neutrino experiments
\cite{hep-ph/0011054,hep-ph/0302039,0709.1937,1104.1390}.

\section{3+2 Neutrino Mixing}

\begin{figure}[t!]
\null
\hfill
\begin{minipage}[l]{0.2\linewidth}
\begin{center}
\includegraphics*[bb=224 471 342 778, width=0.9\linewidth]{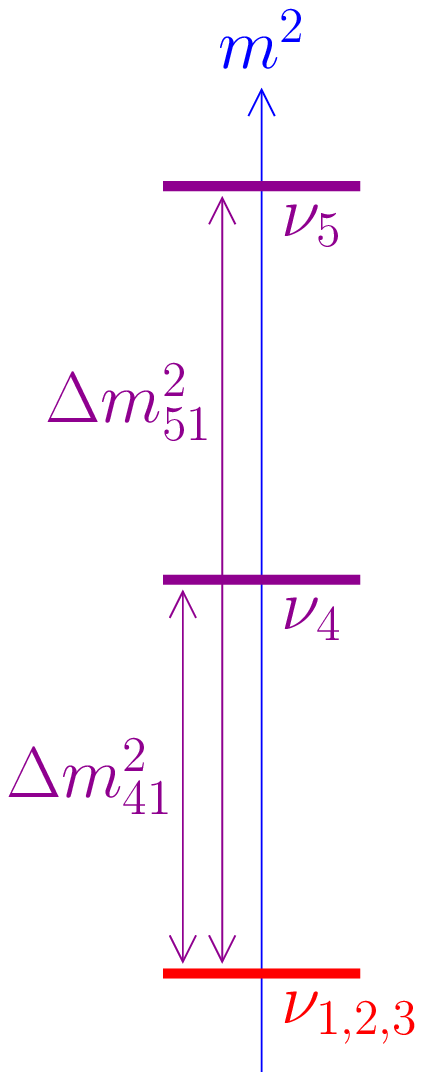}
\\%[0.3cm]
\text{{"normal"}}
\end{center}
\end{minipage}
\hfill
\begin{minipage}[l]{0.2\linewidth}
\begin{center}
\includegraphics*[bb=224 471 342 778, width=0.9\linewidth]{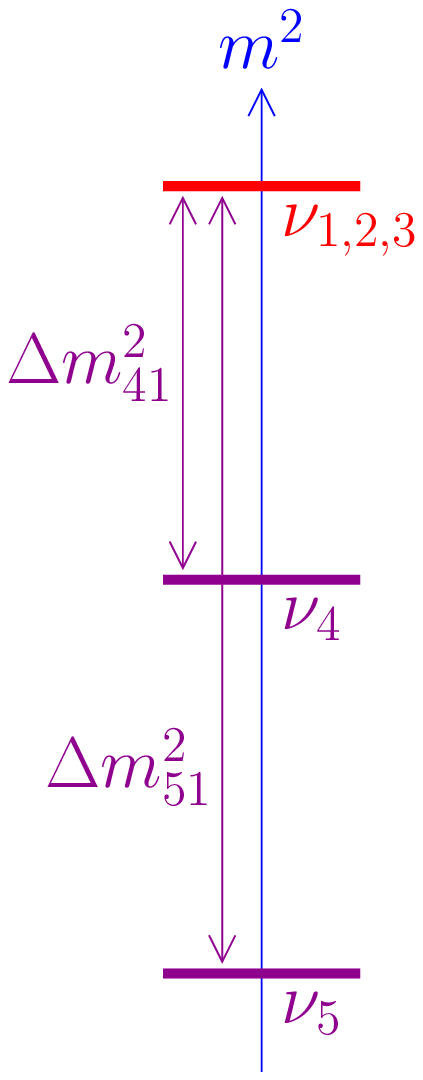}
\\%[0.3cm]
\text{{"inverted"}}
\end{center}
\end{minipage}
\hfill
\begin{minipage}[l]{0.2\linewidth}
\begin{center}
\includegraphics*[bb=224 471 342 778, width=0.9\linewidth]{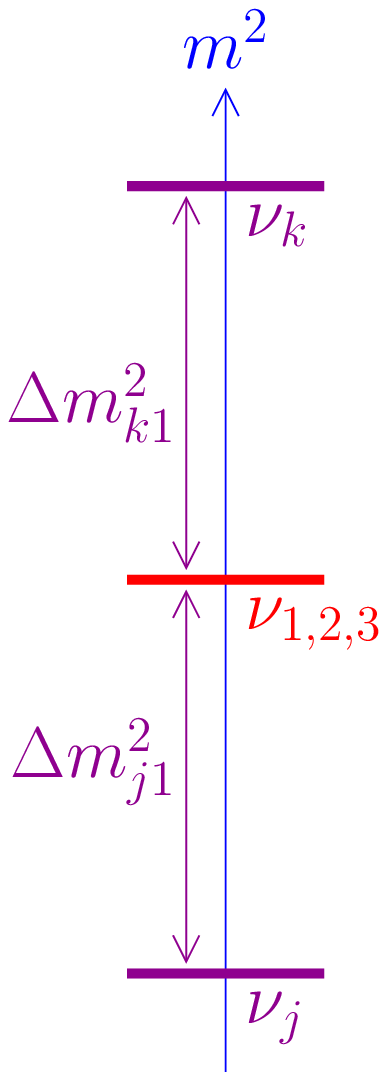}
\\%[0.3cm]
\text{{"perverted"}}
\end{center}
\end{minipage}
\hfill
\null
\caption{ \label{3+2}
Schematic description of the three possible
3+2 schemes that we are considering, taking into account that
$|\Delta{m}^2_{21}| \ll |\Delta{m}^2_{31}| \ll |\Delta{m}^2_{41}| < |\Delta{m}^2_{51}|$.
In the perverted scheme
the identification of the labels $k$ and $j$ is chosen in order to satisfy the inequality
$|\Delta{m}^2_{41}| < |\Delta{m}^2_{51}|$.
}
\end{figure}

The CP-violating difference between MiniBooNE neutrino and antineutrino data
can be explained by introducing another physical effect in addition to a sterile neutrino:
a second sterile neutrino in 3+2 schemes \cite{hep-ph/0305255,hep-ph/0609177,0906.1997,0705.0107,1007.4171,1103.4570},
non-standard interactions \cite{1007.4171},
CPT violation \cite{1010.1395,1012.0267}.
In this section we discuss the possibility of 3+2 neutrino mixing
according to the possible schemes illustrated schematically in Fig.~\ref{3+2}.
The inverted and perverted schemes have been called, respectively, 2+3 and 1+3+1 in Ref.~\cite{0706.1462}.
Since the
inverted and perverted
schemes have three or four massive neutrinos at the eV scale,
they are disfavored by cosmological data \cite{1006.5276,1102.4774}
over the
normal scheme.
Note also that
all 3+2 schemes are disfavored by the
Big-Bang Nucleosynthesis limit
$N_{\text{eff}}\leq4$ at 95\% C.L.
obtained in Ref.~\cite{1103.1261}.

In 3+2 schemes the relevant effective oscillation probabilities in short-baseline experiments are given by
\begin{align}
\null & \null
P_{\boss{\nu}{\mu}\to\boss{\nu}{e}}^{\text{SBL}}
=
4 |U_{\mu4}|^2 |U_{e4}|^2 \sin^{2}\phi_{41}
\nonumber
\\[-0.3cm]
\null & \null
\phantom{
P_{\boss{\nu}{\mu}\to\boss{\nu}{e}}^{\text{SBL}}
=
}
+
4 |U_{\mu5}|^2 |U_{e5}|^2 \sin^{2}\phi_{51}
\nonumber
\\[-0.5cm]
\null & \null
+
8 |U_{\mu4} U_{e4} U_{\mu5} U_{e5}|
\sin\phi_{41}
\sin\phi_{51}
\cos(\phi_{54} \stackrel{(+)}{-} \eta)
\,,
\label{trans-3p2}
\\[0.3cm]
\null & \null
P_{\boss{\nu}{\alpha}\to\boss{\nu}{\alpha}}^{\text{SBL}}
=
1
-
4 (1 - |U_{\alpha4}|^2 - |U_{\alpha5}|^2)
\nonumber
\\
\null & \null
\phantom{
P_{\boss{\nu}{\alpha}\to\boss{\nu}{\alpha}}^{\text{SBL}}
=
1
-
}
\times
(|U_{\alpha4}|^2 \sin^{2}\phi_{41} + |U_{\alpha5}|^2 \sin^{2}\phi_{51})
\nonumber
\\
\null & \null
\phantom{
P_{\nu_{\alpha}\to\nu_{\alpha}}^{\text{SBL}}
=
}
- 4 |U_{\alpha4}|^2 |U_{\alpha5}|^2 \sin^{2}\phi_{54}
\,,
\label{survi-3p2}
\end{align}
for
$\alpha,\beta=e,\mu$,
with
\begin{equation}
\phi_{kj}
=
\Delta{m}^2_{kj} L / 4 E
\,,
\qquad
\eta
=
\text{arg}[U_{e4}^{*}U_{\mu4}U_{e5}U_{\mu5}^{*}]
\,.
\label{3p2}
\end{equation}
Note the change in sign of the contribution of the CP-violating phase $\eta$
going from neutrinos to antineutrinos,
which allows us
to explain the CP-violating difference between MiniBooNE neutrino and antineutrino data.
In our analysis we consider
$\Delta{m}^2_{41}>0$
and
$\Delta{m}^2_{51}>0$,
with
$\Delta{m}^2_{51}>\Delta{m}^2_{41}$,
which implies
$\Delta{m}^2_{54}>0$.
These assumptions correspond to the normal scheme in Fig.~\ref{3+2},
which is favored by cosmological data,
as noted above.
In any case, the results of our analysis can be applied also to the
inverted scheme
($\Delta{m}^2_{41}<0$,
$\Delta{m}^2_{51}<0$,
$\Delta{m}^2_{54}<0$)
with the change
$\eta \to 2\pi - \eta$.
Instead the perverted schemes,
which have been considered in the fit of Ref.~\cite{1103.4570},
require a separate treatment because in these schemes
$|\Delta{m}^2_{54}| = |\Delta{m}^2_{51}| + |\Delta{m}^2_{41}|$.
For simplicity we do not consider them here,
because they are strongly disfavored by cosmological data,
having four massive neutrinos at the eV scale.

\begin{figure}[t!]
\includegraphics*[width=\linewidth]{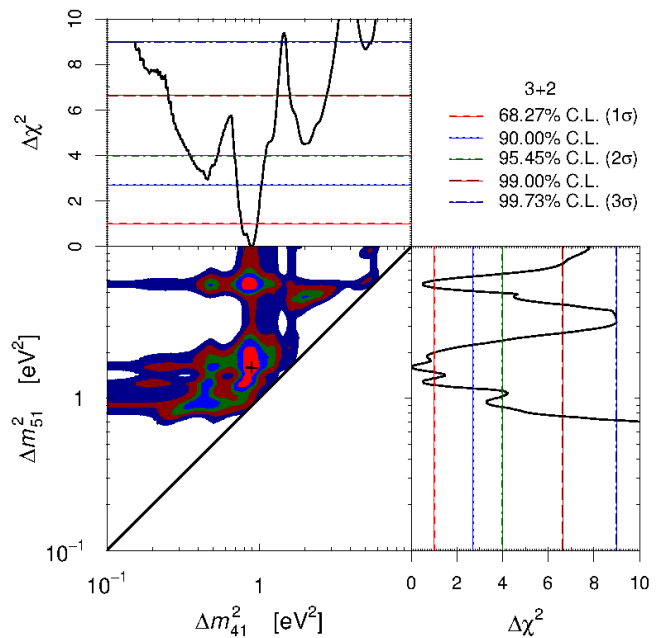}
\caption{ \label{3p2-dm2}
Allowed regions in the
$\Delta{m}^2_{41}$--$\Delta{m}^2_{51}$ plane
and
corresponding
marginal $\Delta\chi^{2}$'s
obtained from the global fit of all the considered data in 3+2 schemes.
The best-fit point corresponding to $\chi^2_{\text{min}}$ is indicated by a cross.
}
\end{figure}

\begin{figure}[t!]
\includegraphics*[width=\linewidth]{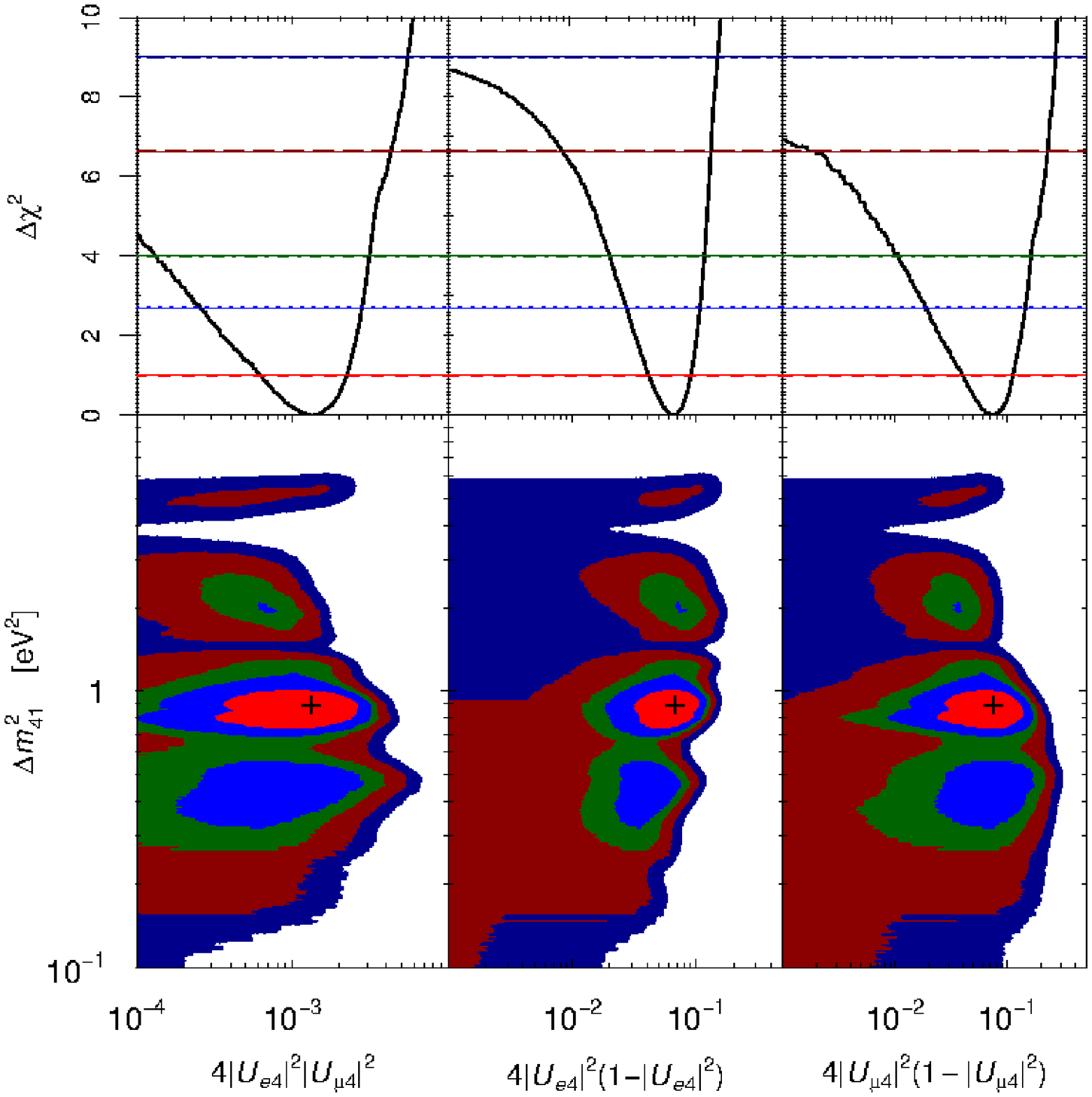}
\caption{ \label{3p2-d41}
Allowed regions in the
$4|U_{e4}|^2|U_{\mu4}|^2$--$\Delta{m}^2_{41}$,
$4|U_{e4}|^2(1-|U_{e4}|^2)$--$\Delta{m}^2_{41}$ and
$4|U_{\mu4}|^2(1-|U_{\mu4}|^2)$--$\Delta{m}^2_{41}$
planes
and
marginal $\Delta\chi^{2}$'s
for
$4|U_{e4}|^2|U_{\mu4}|^2$,
$4|U_{e4}|^2(1-|U_{e4}|^2)$ and
$4|U_{\mu4}|^2(1-|U_{\mu4}|^2)$
obtained from the global fit of all the considered data in 3+2 schemes.
The line types and color have the same meaning as in Fig.~\ref{3p2-dm2}.
The best-fit point corresponding to $\chi^2_{\text{min}}$ is indicated by a cross.
}
\end{figure}

\begin{figure}[t!]
\includegraphics*[width=\linewidth]{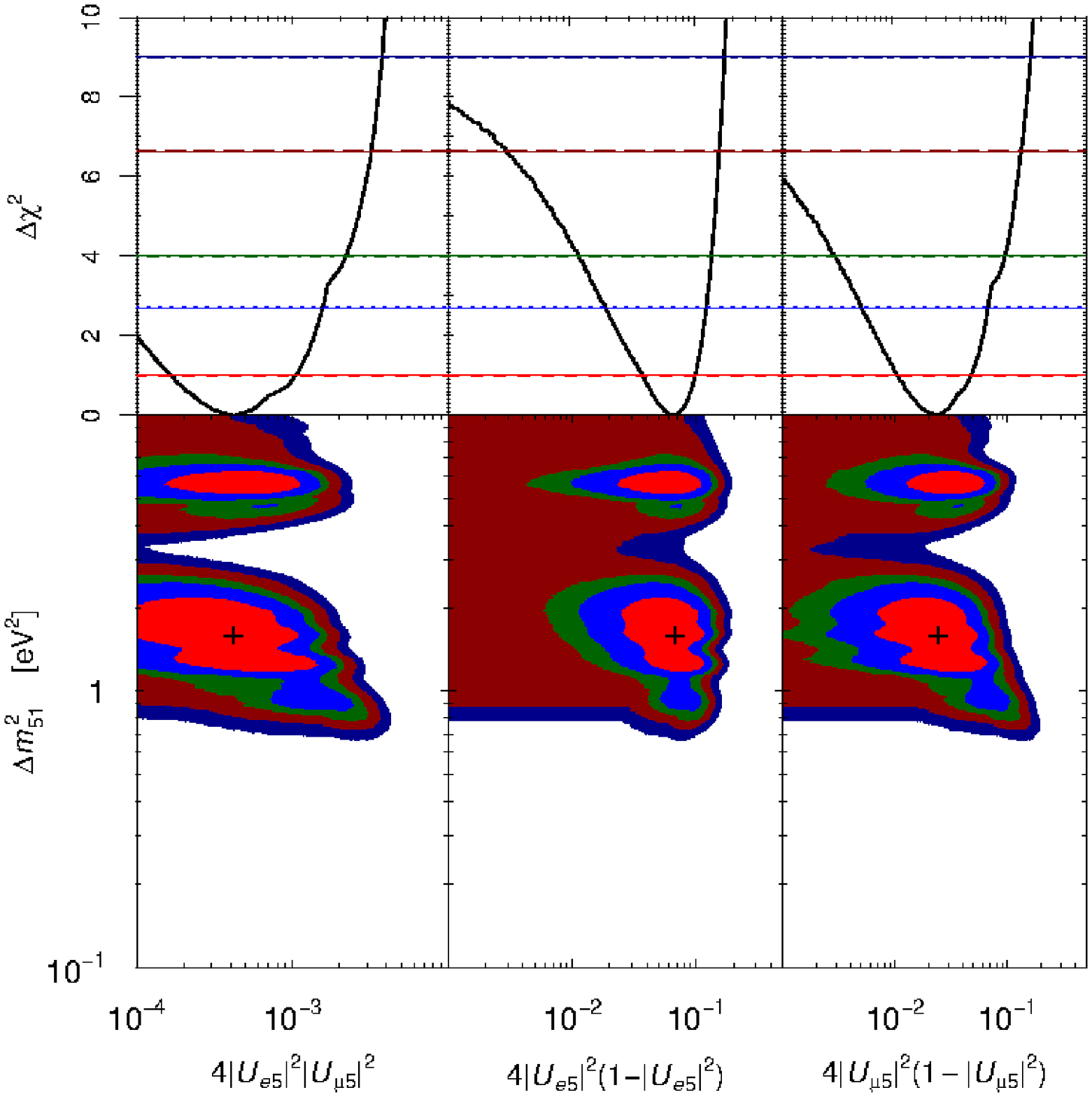}
\caption{ \label{3p2-d51}
Allowed regions in the
$4|U_{e5}|^2|U_{\mu5}|^2$--$\Delta{m}^2_{51}$,
$4|U_{e5}|^2(1-|U_{e5}|^2)$--$\Delta{m}^2_{51}$ and
$4|U_{\mu5}|^2(1-|U_{\mu5}|^2)$--$\Delta{m}^2_{51}$
planes
and
marginal $\Delta\chi^{2}$'s
for
$4|U_{e5}|^2|U_{\mu5}|^2$,
$4|U_{e5}|^2(1-|U_{e5}|^2)$ and
$4|U_{\mu5}|^2(1-|U_{\mu5}|^2)$
obtained from the global fit of all the considered data in 3+2 schemes.
The line types and color have the same meaning as in Fig.~\ref{3p2-dm2}.
The best-fit point corresponding to $\chi^2_{\text{min}}$ is indicated by a cross.
}
\end{figure}

\begin{figure}[t!]
\includegraphics*[width=\linewidth]{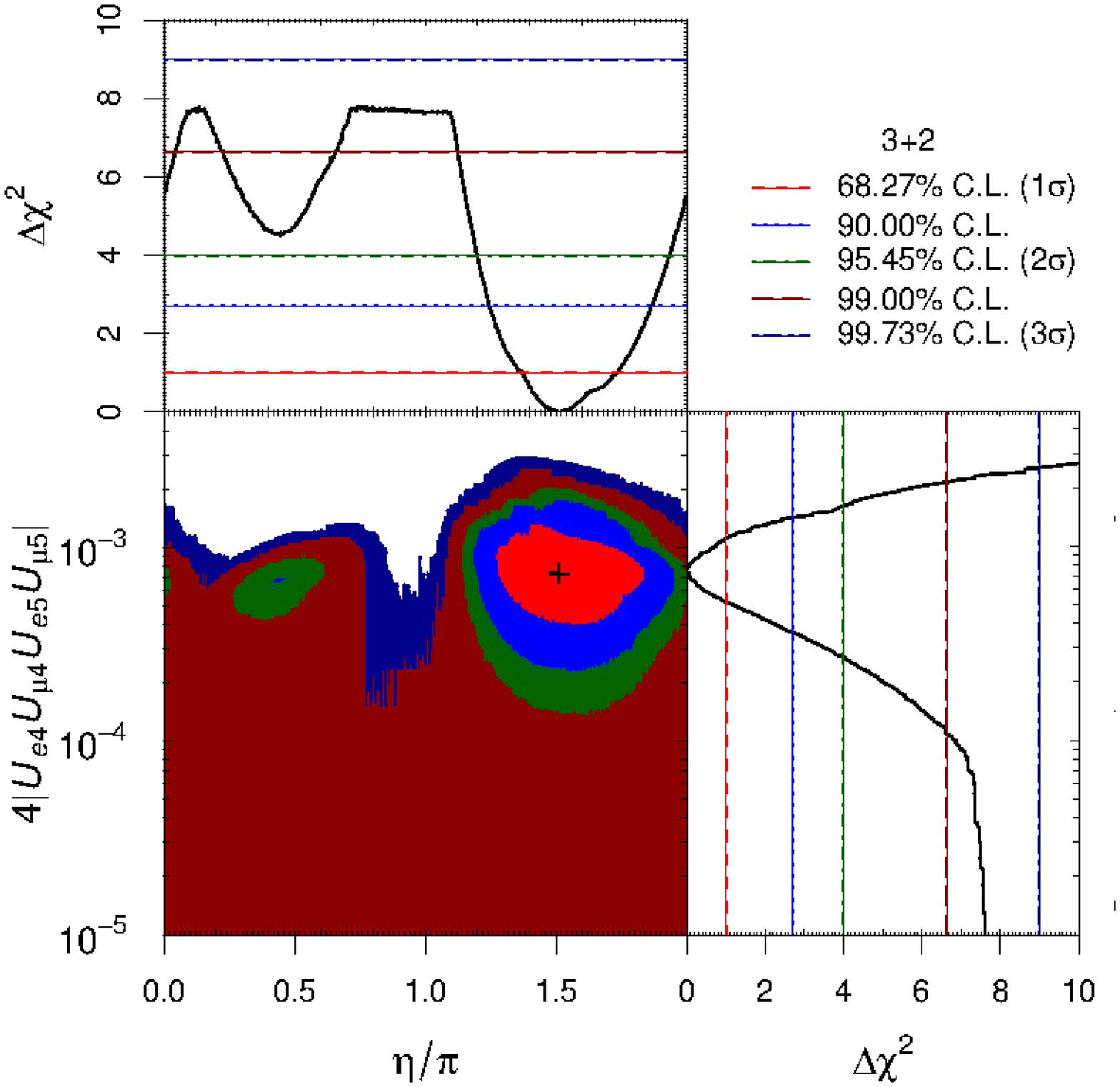}
\caption{ \label{3p2-eta}
Allowed regions in the
$\eta$--$4|U_{e4}U_{\mu4}U_{e5}U_{\mu5}|$
plane
and
corresponding
marginal $\Delta\chi^{2}$'s
obtained from the global fit of all the considered data in 3+2 schemes.
The best-fit point corresponding to $\chi^2_{\text{min}}$ is indicated by a cross.
}
\end{figure}

Figures~\ref{3p2-dm2}--\ref{3p2-eta}
show the marginal allowed regions in two-dimensional planes of interesting combinations of the
oscillation parameters
and the corresponding marginal $\Delta\chi^{2}$'s
obtained in our 3+2 global fit of the same set of data used in Fig.~\ref{exc-nev}.
The best-fit values of the mixing parameters are shown in Tab.~\ref{bef}.

The correlation of the allowed regions of
$\eta$ and $4|U_{e4}U_{\mu4}U_{e5}U_{\mu5}|$
in Fig.~\ref{3p2-eta}
is due to their presence in the last term in the effective
$\boss{\nu}{\mu}\to\boss{\nu}{e}$
oscillation probability in Eq.~(\ref{trans-3p2}).
The marginal $\Delta\chi^{2}$
for $\eta$
has two minima close to the two values where CP-violation is maximal
($\eta=\pi/2$ and $\eta=3\pi/2$),
in agreement with what we expected from the need to fit the
positive $\bar\nu_{\mu}\to\bar\nu_{e}$ signal
and
negative $\nu_{\mu}\to\nu_{e}$ measurement
in the MiniBooNE experiment in the same range of $L/E$.
From Fig.~\ref{3p2-eta} one can also see that
the marginal $\Delta\chi^{2}$ for $\eta$ is always smaller than the
$\Delta\chi^{2} \simeq 7.8$ corresponding to
a negligibly small value of $4|U_{e4}U_{\mu4}U_{e5}U_{\mu5}|$
(this value is reached for $\eta\simeq0.1\pi$ and around $\eta=\pi$).
Such a $\Delta\chi^{2}$ is smaller than the difference of the $\chi^{2}$ minima
in the 3+1 and 3+2 schemes because the condition for
$4|U_{e4}U_{\mu4}U_{e5}U_{\mu5}|$ to vanish requires that only one of
$U_{e5}$ and $U_{\mu5}$ vanishes.
In particular,
if only $U_{\mu5}$ is practically negligible,
the reactor antineutrino data can be fitted sligtly better than in 3+1 schemes,
as already noted in Ref.~\cite{1103.4570}.

The parameter goodness-of-fit obtained with the comparison of the fit of
LSND and MiniBooNE antineutrino data
and the fit of all other data is
$
5
\times
10^{-4}
$.
This is an improvement with respect to the
$
6
\times
10^{-6}
$
parameter goodness-of-fit obtained in 3+1 schemes.
However,
the value of the parameter goodness-of-fit
remains low
and the improvement is mainly due to the increased number of degrees of freedom,
as one can see from Tab.~\ref{bef}.
The persistence of a bad parameter goodness-of-fit
is a consequence of the
fact that the
$\bar\nu_{\mu}\to\bar\nu_{e}$
transitions observed in LSND and MiniBooNE
must correspond in any neutrino mixing schemes to
enough short-baseline disappearance of $\boss{\nu}{e}$ and $\boss{\nu}{\mu}$
which has not been observed
and there is an irreducible tension between the
LSND and MiniBooNE antineutrino data
and the KARMEN antineutrino data.
The only benefit of 3+2 schemes with respect to 3+1 schemes is that
they allow to explain the difference between MiniBooNE neutrino and antineutrino data
through CP violation.
In fact,
neglecting the MiniBooNE neutrino data we obtain
$
\Delta\chi^{2}_{\text{PG}}
=
16.6
$
with
$
\text{PGoF}
=
3
\times
10^{-4}
$
in 3+1 schemes
and
$
\Delta\chi^{2}_{\text{PG}}
=
20.4
$
with
$
\text{PGoF}
=
1
\times
10^{-3}
$
in 3+2 schemes.
In this case
$\Delta\chi^{2}_{\text{PG}}$
is even lower in 3+1 schemes than in 3+2 schemes!

\begin{figure}[t!]
\includegraphics*[width=\linewidth]{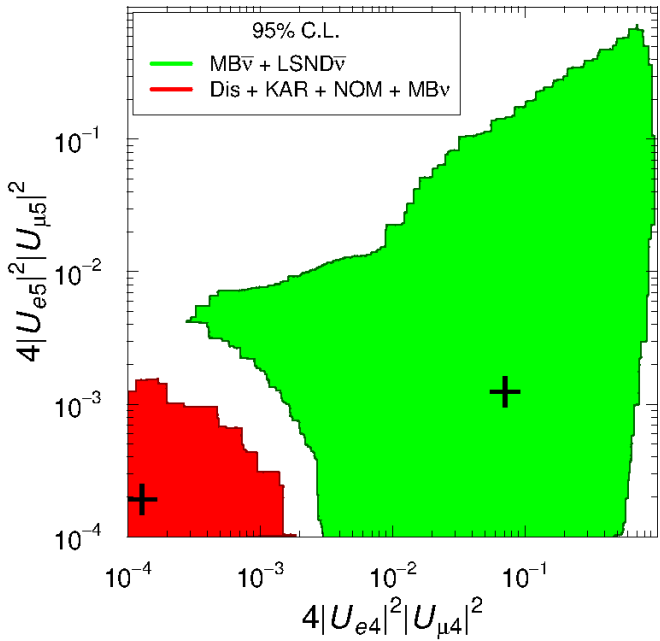}
\caption{ \label{3p2-sup-34}
Comparison of the 95\% C.L. allowed regions in the
$4|U_{e4}|^2|U_{\mu4}|^2$--$4|U_{e5}|^2|U_{\mu5}|^2$ plane
obtained from
LSND and MiniBooNE antineutrino data on the right (green area)
and
disappearance,
KARMEN,
NOMAD
and 
MiniBooNE neutrino
data on the left (red area).
The corresponding best-fit points are indicated by crosses.
}
\end{figure}

The tension between LSND and MiniBooNE antineutrino data
and
disappearance,
KARMEN,
NOMAD
and 
MiniBooNE neutrino
data is illustrated in Fig.~\ref{3p2-sup-34},
which is the analogous for 3+2 schemes of Fig.~\ref{exc-nev} in 3+1 schemes.
In practice, in order to show the tension in a two-dimensional figure we have marginalized the $\chi^2$ over all the other mixing parameters,
including the two $\Delta{m}^2$'s.

The results of our 3+2 global fit are in reasonable agreement with those presented in
Ref.~\cite{1103.4570}.
There is a discrepancy in the location of the best-fit point in the
$\Delta{m}^2_{41}$--$\Delta{m}^2_{51}$
plane,
but we obtain similar regions for the local $\chi^2$ minima.
Our allowed regions are larger than those presented in
Ref.~\cite{1103.4570}.
We think that such difference is probably due to a different treatment of the
spectral data of the Bugey-3 reactor experiment \cite{Declais:1995su}
which cause the wiggling
for
$\Delta{m}^2 \lesssim 1 \, \text{eV}^2$
of the disappearance limit
in Fig.~\ref{exc-dis}
and the exclusion curve
in Fig.~\ref{exc-nev}.
Such wiggling is wider in Fig.~3 of Ref.~\cite{1103.4570},
leading to deeper valleys of the $\chi^2$ function
and
smaller allowed regions.
The compatibility with cosmological data of the allowed regions in the
$\Delta{m}^2_{41}$--$\Delta{m}^2_{51}$
plane shown in Fig.~\ref{3p2-dm2}
will be discussed in a separate article
\cite{Archidiacono-Fornengo-Giunti-Melchiorri-IP-11}
(an interesting previous study was presented in Ref.~\cite{0810.5133}).

Figures~\ref{3p2-d41}--\ref{3p2-d51}
show the allowed regions for the amplitudes of the oscillating terms in short-baseline
$\boss{\nu}{\mu}\to\boss{\nu}{e}$ transitions
and
$\boss{\nu}{e}$
and
$\boss{\nu}{\mu}$
disappearance, for which we have the best-fit values
\begin{align}
\null & \null
4|U_{e4}|^2|U_{\mu4}|^2
=
0.0013
\,,
\label{s44}
\\
\null & \null
4|U_{e5}|^2|U_{\mu5}|^2
=
0.00042
\,,
\label{s55}
\\
\null & \null
4|U_{e4}|^2(1-|U_{e4}|^2)
=
0.068
\,,
\label{se4}
\\
\null & \null
4|U_{e5}|^2(1-|U_{e5}|^2)
=
0.068
\,,
\label{se5}
\\
\null & \null
4|U_{\mu4}|^2(1-|U_{\mu4}|^2)
=
0.076
\,,
\label{sm4}
\\
\null & \null
4|U_{\mu5}|^2(1-|U_{\mu5}|^2)
=
0.024
\,.
\label{sm5}
\end{align}
Comparing the values of
$4|U_{e4}|^2|U_{\mu4}|^2$,
$4|U_{e4}|^2(1-|U_{e4}|^2)$ and
$4|U_{\mu4}|^2(1-|U_{\mu4}|^2)$
with those obtained in 3+1 mixing, given in Eqs~(\ref{sem})--(\ref{smm}),
one can see that they are lower,
but keep the same order of magnitude.
In the fit of the data
the smaller values of these amplitudes is due to the additional contribution of the amplitudes
generated by the mixing of
$\nu_{e}$ and $\nu_{\mu}$
with $\nu_{5}$.

\section{Conclusions}

In this paper we presented the results of fits of short-baseline neutrino oscillation data
in 3+1 and 3+2 neutrino mixing schemes.

In the framework of 3+1 neutrino mixing schemes in Fig.~\ref{3+1},
we confirm the strong tension between LSND and MiniBooNE antineutrino data
and
disappearance,
KARMEN,
NOMAD
and 
MiniBooNE neutrino
data
discussed recently in Refs.~\cite{0705.0107,1007.4171,1012.0267,1103.4570}.
Since however the minimum value of the global $\chi^2$ is rather good,
one may choose to consider as possible 3+1 neutrino mixing,
which can partially explain the data,
taking into account its simplicity and the natural correspondence of one new entity
(a sterile neutrino)
with a new effect
(short-baseline oscillations).
Following this approach, we presented the results of the global fit in 3+1 neutrino mixing,
which leads to the determination of restricted allowed regions in the mixing parameter space
which can be explored in future
$\boss{\nu}{\mu}\to\boss{\nu}{e}$
\cite{CRUBBIA2011,1007.3228,1105.4984}
$\boss{\nu}{e}$
disappearance
\cite{0907.3145,0907.5487,1006.2103,Porta:2010zz,1011.4509,1103.5307,CRUBBIA2011,Pallavicini2011,1105.4984}
and
$\boss{\nu}{\mu}$
disappearance
\cite{CRUBBIA2011,1106.5685}
experiments.

We also presented a global fit
in the framework of the 3+2 neutrino mixing schemes in Fig.~\ref{3+2}.
We have shown that the tension between LSND and MiniBooNE antineutrino data
and
disappearance,
KARMEN,
NOMAD
and 
MiniBooNE neutrino
data
is reduced with respect to the 3+1 fit,
but is not eliminated
(see Fig.~\ref{3p2-sup-34}).
Moreover,
the improvement of the parameter goodness of fit with respect to that obtained in the 3+1 fit
is mainly due to the increase of the number of oscillation parameters,
as one can see from Tab.~\ref{bef}.
Hence it seems mainly a statistical effect.

The results of our 3+2 fit are compatible with those presented recently in
Ref.~\cite{1103.4570},
but we obtain a different indication for the best fit
(see Tab.~\ref{bef}).
%(see Eqs.~(\ref{3p2-bef1})--(\ref{3p2-bef4})).
For the CP-violating phase we obtained
two minima of the marginal $\chi^2$ close to the two values where CP-violation is maximal.

In conclusion,
we think that our results are useful for the discussion of the interpretation of
the current experimental indications in favor of short-baseline neutrino oscillations
and for the study of new experiments aimed at
a clarification of the validity of these indications.

\bibliography{bibtex/nu}

\end{document}